\documentclass[12pt]{article}
\usepackage{amsmath} \usepackage{amsfonts}
\usepackage{amsxtra}
\usepackage{amssymb}
\usepackage[english]{babel}
\usepackage{epsfig}

\newcounter{apps}

\newcounter{prs}[section]

\newcounter{cors}

\newcounter{th}[section]

\newcounter{lm}[subsection]

\newcounter{df}[section]

\newcommand{\be}{\begin{equation}}
\newcommand{\ee}{\end{equation}}
\newcommand{\bea}{\begin{eqnarray*}}
\newcommand{\eea}{\end{eqnarray*}}
\newcommand{\beaa}{\begin{eqnarray}}
\newcommand{\eeaa}{\end{eqnarray}}

\newcommand{\ba}{\begin{array}}
\newcommand{\ea}{\end{array}}

\newcommand{\lb}{\label}
\newcommand{\ra}{\rightarrow}

\newcommand{\wt}{\widetilde}
\newcommand{\td}{\tilde}

\newcommand{\g}{\gamma}
\newcommand{\G}{\Gamma}

\newcommand{\bt}{\beta}
\newcommand{\p}{\partial}

\newcommand{\Ld}{\Lambda}
\newcommand{\vp}{\varphi}

\newcommand{\RR}{{\mathbb{R}}}

\newcommand{\Hd}{{\mathcal{H}}^*}
\newcommand{\HH}{{\mathcal{H}}}

\begin{document}

\begin{flushright}
ITEP-TH-27/04\\
\end{flushright}

\vspace{1.0cm}

\begin{center}

{\large\bf
Hopf algebra of graphs and the RG equations
}

\bigskip

D.V.Malyshev
\footnote{
dmalyshe@princeton.edu}

\bigskip


\end{center}

\bigskip
\bigskip

\bigskip

\begin{abstract}

We study
the renormalization group equations following from the Hopf
algebra of graphs.
Vertex functions are treated as vectors
in dual to the Hopf algebra space.
The RG equations on such vertex functions are equivalent to
RG equations on individual Feynman integrals.
The solution to the RG equations may be represented as an exponent of
the beta-function.
We explicitly show that the exponent of the one-loop beta function enables
one to find the coefficients in front of the
leading logarithms for individual Feynman integrals.
The same results are obtained in parquet approximation.

\end{abstract}


\section{Generalized RG equation}

One loop RG equations enable one to find the leading logarithm for the
sum of Feynman diagrams of a given order in coupling constant
\cite{BSh,Col}.
In fact there exists an RG equation for individual Feynman integrals
\cite{chetyr:ll}
and the one loop beta-function
enables one to find the power and the coefficient in front of the
leading logarithm for any Feynman integral \cite{mllrg,mllnrg}.
The number of known coefficient of the beta-function is equal to the
number of sub-leading logarithms one can find.

In the paper it is shown that the RG equation for individual Feynman
diagrams follows from the RG equation dual to the Hopf algebra of
graphs.
The Hopf algebra of graphs was introduced by Connes and Kreimer as a
mathematical structure underlying the Bogolubov R-operation \cite{CK1}.
In this formalism, Feynman integrals are elements of the dual space of
linear functions on graphs.
The Hopf algebra of graphs induces some Lie algebra in the dual space.
The beta-function is an element of the Lie algebra.
Exponentiation of the beta-function defines a diffeomorphism in the
space of Feynman integrals%
\footnote{Strictly speaking, Feynman integrals are characters of the
Hopf algebra, thus the diffeomorphism should be a diffeomorphism of
the subspace of characters.
}.
In the paper it is shown explicitly that the exponent of the one-loop
beta-functions gives the leading logarithms for individual Feynman
integrals.

We shall study the massless $\vp^4$ theory in four dimensional
euclidian space with the lagrangian
$$
L=\frac{1}{2}(\p_\mu\vp)^2-\frac{16\pi^2}{4!}g\vp^4.
$$
The minus sign in interaction is introduced in order to avoid the
minus signs in the perturbation expansion.

Further we discuss the four point function
\be\lb{vf-class}
F=\sum_{n=0}^\infty F_n g^{n+1},
\ee
where $F_n$ is the contribution of $n$-loop diagrams.

Vertex functions in perturbation theory are formal series in the coupling
constant.
The space of formal series is an infinite dimensional space where
the monomials
$g^n$ form the basis.
Analogously one can consider the linear space of graphs, i.e.
an infinite dimensional space where the basis vectors are labeled by
graphs.
We will work in the space where the basis vectors correspond to one
particle irreducible graphs (1PI graphs), to disjoint unions of 1PI
graphs, and also there is a vector corresponding to empty graph
$\emptyset$.
Denote the linear space of graphs by $\HH$.
The graphs and the corresponding vectors in $\HH$
will be denoted by $\G$ and $\g$.

Vertex functions belong to the space $\Hd$ of linear functions on
graphs.
In the dual basis the four-point vertex function has the form
\be\lb{vf-new}
F=\g_0+\sum_\g \frac{1}{S_\g}F_\g\g,
\ee
where
$\g_0$ is the graph consisting of one vertex,
the graphs $\g$ are the 1PI graphs with four external edges,
$F_\g$ is the value of Feynman integral for the graph $\g$,
$S_\g$ is the symmetry factor of $\g$.
Symbol $\g$ on the right denotes the basis vector
in $\Hd$.

The linear space of graphs is an extension
of the space of formal series.
If one substitutes graphs with $k$ vertices by $g^{k}$,
then the vertex function
(\ref{vf-new}) becomes a usual vertex function and the RG equations
become the standard RG equations.
But after this substitution some information will be lost.

The definitions above are distinct from the definitions of Connes and
Kreimer.
In the work \cite{CK2}
the effective coupling is considered to be a formal series in coupling
constant with coefficients in the algebra of graphs, i.e. in formula
(\ref{vf-class}) the coefficient $F_n$ becomes an element of $\HH$.
Whereas in our definition the space of formal series is substituted by
the space of linear functions $\Hd$, i.e. in formula
(\ref{vf-class})  $g^n$ is substituted by vectors from $\Hd$.

Another distinction is that we fix the external edges.
This means, for example, that in vertex function (\ref{vf-new})
we have three one loop graphs with different orientations
$\g_1^{(s)}$, $\g_1^{(t)}$ and $\g_1^{(u)}$.
The coefficients in front of these graphs are
$F(\g_1^{(s)})=\log\frac{\mu^2}{s}$,
$F(\g_1^{(t)})=\log\frac{\mu^2}{t}$ and
$F(\g_1^{(u)})=\log\frac{\mu^2}{u}$.

One can find the RG equations in the linear space of graphs using the
analogy with the RG equation in the space of formal series in coupling
constant.
The RG equation on the four-point function has the form
\be\lb{genurg}
\frac{d}{d\log\mu^2}Z^{-2}F=0,
\ee
where $Z$ is the renormalization of the fields.
In massless theory in dimensional regularization the mass does not
appear in any order of perturbation.
In one loop approximation there are no renormalizations of the
two-point vertex function, i.e. there are no renormalizations of the
fields.
Consequently in one loop approximation equation (\ref{genurg})
has a more simple form
\be\lb{1lurg}
\frac{d}{d\tau}F=0,
\ee
where we denote $\tau=\log\mu^2$.
Equation (\ref{1lurg}) may be written in the form of
Callan-Symanzik equation
\be\lb{c-sym}
\p_\tau F+\bt(g)\p_g F=0.
\ee
Where the derivative $\p_\tau$ acts on Feynman integrals $F_n$
and the operator $\bt(g)\p_g$ acts on $g^n$, i.e. on the basis vectors
in the space of formal series
\be\lb{beta-act}
\bt(g)\p_g g^{n}=-\frac{3}{2}ng^{n+1}.
\ee
Using (\ref{beta-act}), we can rewrite equation (\ref{c-sym})
in the form
\be
\sum_{n=0}^\infty (\frac{d}{d\tau}F_n-\frac{3}{2}nF_{n-1})g^{n+1}=0.
\ee
This equation is equivalent to the system of equations
\be
\frac{d}{d\tau}F_n -\frac{3}{2}nF_{n-1}=0,\;\;\;n=0,1,2\ldots
\ee
The solution of the system gives the leading logarithm for the sum
of $n$-loop diagrams
\be\lb{ll2}
F_n=\left(\frac{3}{2}\tau\right)^n,
\ee
where $\tau=\log\frac{\mu^2}{p^2}$.

In order to find the generalization of equation
(\ref{c-sym}) in the linear space of graphs
we have to define the action of the beta-function on basis vectors
corresponding to graphs.
From the point of view of the Hopf algebra $\HH$,
beta-function is an element of the dual space $\Hd$
where the multiplication is defined as an insertion of graphs
\cite{CK1}
\be
\g_1*\g_2=\sum_v\g_1\circ_v\g_2+\g_1\cdot\g_2,
\ee
symbol $\circ_v$ denotes the insertion of the graph $\g_1$
in the vertex $v\in\g_2$
and $\g_1\cdot\g_2$ denotes the
disjoint union of $\g_1$ and $\g_2$.

The generalization of equation (\ref{c-sym}) has the form
\be\lb{KSHH}
\p_\tau F+\hat{\bt}\circ F=0,
\ee
here $F$ is a vector in $\Hd$
that we define in (\ref{vf-new}) as a sum over 1PI graphs.
Operator $\hat{\bt}$ is the one-loop beta-function
\be
\hat{\bt}=-\frac{1}{2}(\g_1^{(s)}+\g_1^{(t)}+\g_1^{(u)}).
\ee
The derivative $\p_\tau$ acts on the coordinates, i.e. on Feynman
integrals $F_\g$.
The operator $\hat{\bt}$ acts on basis vectors
$\g$ by insertion of one-loop diagrams.
Note, that the analogous action of the operator
$\bt(g)\p_g$ is the insertion of $\bt(g)$ instead of $g$.

Equation (\ref{KSHH}) is an equation on vector $F$.
It is equivalent to a system of equations on the coordinates of this
vector
\be\lb{rgmain}
\frac{\p}{\p\tau}F(\g_{n+1})=\sum_{\g_{n}=\g_{n+1}/\g_1}F(\g_{n}),
\ee
where  $F(\g_{n+1})$ is the leading logarithm for the $(n+1)$-loop
graph $\g_{n+1}$.
The sum is over $n$-loop graphs $\g_{n}$ obtained after contraction of
one-loop subgraphs in $\g_{n+1}$.
Symmetry factors $S_{\g}$ in (\ref{rgmain}) vanish due to the
equality \cite{mllrg}
\be
\frac{i(\g_1,\g_n;\g_{n+1})}{S_{\g_1}S_{\g_n}}=\frac{1}{S_{\g_{n+1}}},
\ee
where $i(\g_1,\g_n;\g_{n+1})$ is the number of orientations of $\g_1$
such that the insertion of $\g_1$ in $\g_n$ yields $\g_{n+1}$.

Note, that formula (\ref{rgmain}) incorporates only leading
logarithms.
In order to find sub-leading logarithms one has to contract
sub-diagrams with corresponding number of loops and multiply by the
coefficients of beta-function for these sub-diagrams
\cite{chetyr:ll}.

\section{Diffeomorphisms and shifts}

In this section we find the coefficients in front of the leading
logarithms for individual Feynman integrals using the solution of
one-loop RG equation (\ref{KSHH}) defined in the space $\Hd$.
This equation has the form of Schr\"{o}dinger equation in euclidian space
\be\lb{Sch-HH}
\p_\tau F=-\hat{\bt}\circ F,
\ee
where the beta-function plays the role of hamiltonian.
The solution of this equation is
\be
F(\tau)=e^{-\tau \hat{\bt}}F_0,
\ee
where $F_0$ is an initial vertex function.
The operator $e^{-\tau \hat{\bt}}$ describes the evolution of
initial vector $F_0$ to the vector $F(\tau)$.
In other words, this operator defines a diffeomorphism of the space
$\Hd$ corresponding to the RG transformation.
The problem is that the action of
$e^{-\tau \hat{\bt}}$ has a complicated structure.
The operator $\hat{\bt}$ acts not only on functions but
also on the other operators $\hat{\bt}$ in the exponent.
In order to understand the structure of the answer let us consider a
more simple problem about the diffeomorphisms of the real line.

Let $\hat{V}=V(x)\p_x$ be a vector field,
then the operator
$$
g=e^{\hat{V}}
$$
defines a diffeomorphism of $\RR^1$.
This operator is invariant with respect to the changes
$x\ra y(x)$, but the action of this operator on functions $f(x)$
is quite complicated since the differentiation acts not only on
functions but also on $V(x)$.
It is more convenient to work with normal ordered operators.
The normal ordering is denoted by double dots on both sides of the
operator
\be
\wt{g}=:e^{\wt{V}(x)\p_x}:
\ee
and means that the derivative acts on the function and
doesn't act on $\wt{V}(x)$ in the exponent, consequently
\be
:e^{\wt{V}(x)\p_x}:f(x)=f(x+\wt{V}(x)).
\ee
Thus the normal ordered operator defines a shift of the variable.
The problem is to find for a given $V(x)$ corresponding $\wt{V}(x)$
such that
\be
e^{V(x)\p_x}=:e^{\wt{V}(x)\p_x}:
\ee
The beginning of expansion of $e^{V\p}$ is
\be
e^{V(x)\p_x}=1+V\p+\frac{1}{2}((V\p V)\p+V^2\p^2)+\ldots
\ee
The beginning of expansion of $\wt{V}(x)$ is then
\be
\wt{V}(x)=V+\frac{1}{2}V\p V+\frac{1}{3!}(V\p V\p V+V^2\p^2 V)+ \ldots
\ee
In general we have \cite{CM,GMS}
\be\lb{norup}
\wt{V}(x)=\sum_T\frac{1}{S_T}\frac{1}{T^!}V_T,
\ee
where the sum is over rooted trees $T$, $S_T$ is the number of symmetries
of the rooted tree $T$, i.e. the number of permutations of the edges
of the tree that don't change the tree.
In order to define the tree factorial $T^!$
let us draw the rooted tree $T$ so that the root will be on the top
and all the edges will look downwards.
Then given a vertex one can define a subtree as the part of the tree
below this vertex.
Denote by $n_t$ the number of the vertices in the subtree
$t\subset T$.
The tree factorial is
\be
T^!=\prod_{t\subseteq T}n_t,
\ee
where the product is over all subtrees in $T$.
The ordinary factorial $N!$ equals the tree factorial for the tree
with $N$ vertices and without branching.

The last step is to define the function $V_T$.
This function depends on $V$ and its derivatives.
In order to define $V_T$
it is helpful to consider fields depending on several variables.
Then
\be\lb{ds}
\wt{V}^i=V^i+\frac{1}{2}V^j\p_j V^i
+\frac{1}{3!}(V^k\p_k V^j\p_j  V^i+V^k V^j\p_k\p_j  V^i)+ \ldots
\ee
Functions $V^i(x)$ correspond to vertices of the tree,
derivatives $\frac{\p}{\p x^i}$ correspond to the edges.
The root of the tree corresponds to the field with free index $V^i$,
the root has several outgoing edges: they correspond to the derivatives
acting on $V^i$.
Let an edge $P$ correspond to a derivative $\p_k$,
then the vertex on the lower end of $P$ corresponds to the function
$V^k$ with the same index.
The edges that go down from this vertex correspond to the
derivatives acting on $V^k$ and so on.
Consequently for any rooted tree one can find the corresponding term
$V_T$ and inversely for any term $V_T$ in the expansion of
$\wt{V}$ there exists a rooted tree.

Consider the first order differential equation analogous to
(\ref{Sch-HH})
\be
\p_\tau f(x,\tau)=V(x)\p_x f(x,\tau).
\ee
The solution of this equation is
\be
f(x,\tau)=e^{\tau V(x)\p_x}f_0(x)
\ee
where $f_0(x)$ is the initial condition.
The solution may be reexpressed in terms of the normal ordered
operators
\be
f(x,\tau)=:e^{\wt{V}(x,\tau)\p_x}:f_0(x)
=f_0(x+\wt{V}(x,\tau)).
\ee
Let $f_0(x)=x$, then
\be
f(x,\tau)=x+\wt{V}(x,\tau)
=x+\sum_T\frac{1}{S_T}\frac{1}{T^!}\tau^{n_T}V_T,
\ee
where $n_T$ is the number of vertices of $T$.

Let us find the correspondence between  $f(x,\tau)$ and the vertex
function.
The variable $x$ corresponds to the vertex of the graph.
For example,
the variable $x$ along corresponds to the graph $\g_0$
consisting of one vertex.
The operator $V\p$ replaces $x$ by $V(x)$
and corresponds to the beta-function which
replaces the vertex by graphs.
The function $V_T$ is constructed by insertion of $V$ in place of $x$.
Whereas the graphs are obtained by insertion of one-loop graphs
in place of vertices.
The difference is that a graph may correspond to different trees,
i.e. there may exist different ways of insertion that give the same
graph.
Also the symmetry factors of the trees corresponding to graphs are
equal to the symmetry factors of the graphs.
Then the expression for the vertex function in leading logarithmic
approximation is
\be\lb{sol-vf}
F=e^{-\tau \hat{\bt}}\g_0
=\g_0+\wt{\bt}(\tau),
\ee
where
\be\lb{sol-beta}
\wt{\bt}(\tau)
=\sum_\g \frac{1}{S_\g}\sum_{T_\g}\frac{1}{T_\g^!}\tau^{n_\g}\g,
\ee
Now it is easy to find the leading logarithms for individual Feynman
integrals,
using the definition (\ref{vf-new}) of the vertex function,
\be\lb{vf-new1}
F=\g_0+\sum_\g \frac{1}{S_\g}F_\g\g.
\ee
The leading logarithm for an  $n$-loop graph $\g_n$ has the form
\be
F_{\g_n}=c(\g_n)\tau^n.
\ee
The coefficient may be found from equations
(\ref{sol-beta}) and (\ref{vf-new1})
\be\lb{coef1}
c(\g_n)=\sum_{T_{\g_n}}\frac{1}{T_{\g_n}^!}.
\ee
This coefficient should satisfy
the recursive relation following from equation (\ref{rgmain})
\be\lb{rere}
c(\g_{n+1})=\frac{1}{n+1}\sum_{\g_{n}=\g_{n+1}/\g_1}c(\g_{n}),
\ee
One can easily prove
that the coefficient in (\ref{coef1}) satisfies (\ref{rere})
using the following property of
tree factorials
\cite{Kreimer:treef1}
\be\lb{comtr}
\frac{n}{T^!}=\sum_{t\in{\mathcal{F}}(T)}\frac{1}{t^!}\;,
\ee
where $T$ is a rooted tree with $n$ vertices,
${\mathcal{F}}(T)$ is the set of trees $t$ with $(n-1)$
vertices obtained from $T$ by cutting a vertex on the ends of branches.
Note that vertices on the ends of branches correspond to one-loop
subgraphs, consequently the contraction of a one-loop subgraph corresponds
to the cutting of a vertex on the end of a branch.

We see that the leading logarithms for individual Feynman integrals
may be found without calculation of the integrals.
The logic of the derivation is the following:
write the one-loop RG equation in the space $\Hd$;
solve the equation as an exponent of the beta-function;
find the corresponding shift operator, i.e. the normal ordered
exponent;
then the answer is equal to the initial function with shifted argument.
In quantum field theory this shift corresponds to the renormalization of
coupling.

Transition to the ordinary vertex function and the RG equation is provided by
substitution $\g_n\ra g^{n+1}$ and
$\hat{\bt}\ra\bt(g)\p_g$.
The renormalized vertex function is then
\bea\lb{}
F(g,\tau)&=&e^{-\tau \bt(g)\p_g}g\\
&=&g+\wt{\bt}(\tau)\\
&=&g+\sum_{\g\neq\g_0}
\frac{1}{S_\g}\sum_{T_\g}\frac{1}{T_\g^!}\tau^{n_\g}g^{n_\g+1}\\
&=&\frac{g}{1-\frac{3}{2}g\tau},
\eea
where
$\tau=\ln\frac{\Ld^2}{p^2}$ and $g=g(\Ld)$ is the running coupling
defined on the scale $\Ld$.

\section{Parquet approximation}

The problem of the leading logarithmic asymptotics for
the four-point function with arbitrary external momenta
in massless $\vp^4$ theory
was solved by A.M.Polyakov with the help of an integral equation on
the vertex function \cite{polyakov}.
Analogues logic may be applied in order to find the asymptotics of
individual Feynman integrals.

In this section we will rederive formula (\ref{coef1})
for the leading logarithms in symmetric point using the direct
estimations of Feynman integrals in cut off regularization.
Before discussing the general case we consider several examples.

\begin{figure}[h]
\begin{center}
\leavevmode
\epsfxsize 150pt
\epsfig{file=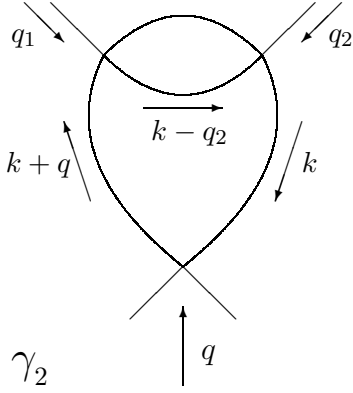}
\caption{\lb{2loop}\bf Two-loop diagram}
\end{center}
\end{figure}

Let $\g_2$ be the diagram shown in figure 1. 
Here $k$ is the loop momentum for the big loop and $p$ is the loop
momentum for the small loop
(this momentum is not shown in the figure).
The momentum $k-q_2$ is external for the one-loop subdiagram.
Let $q\sim q_1\sim q_2$.
Later we will see that the main contribution comes from the region
$k>q$.
Consequently for the internal integral that corresponds to the
one-loop subdiagram we have
\be
F_{\g_1}(k)=\int_{0}^{\Ld}
d^{4}p\:\frac{1}{p^{2}}\:\frac{1}{(p+k)^{2}},
\ee
The leading logarithmic contribution for this integral comes from the
region $p>k$, consequently the lower limit may be replaced by $k$
and the integrand may be replaced by
$d\ln\frac{p^2}{\Ld^2}$.
Thus in the leading logarithmic approximation the integral is
\be
F_{\g_1}(p)=\int_{k}^{\Ld}d\ln\frac{p^2}{\Ld^2}
=\int_0^y dx=y,
\ee
where $y=\ln\frac{\Ld^2}{k^2}$ and $x=\ln\frac{\Ld^2}{p^2}$.
Now consider the integral in momentum $k$ for the diagram $\g_2$,
using the answer for the one-loop subdiagram,
\be
F(\g_2)
=\int_{q}^{\Ld}
d^{4}k\:\frac{1}{k^{2}}\:\frac{1}{(k+q)^{2}}
\ln\frac{\Ld^2}{k^2}
\ee
The logarithm comes from the region $k>q$, thus
\be
F(\g_2)
=\int_0^\eta y dy
=\frac{1}{2}\eta^2,
\ee
where $\eta=\ln\frac{\Ld^2}{q^2}$.

\begin{figure}[h]
\begin{center}
\leavevmode
\epsfxsize 350pt
\epsfig{file=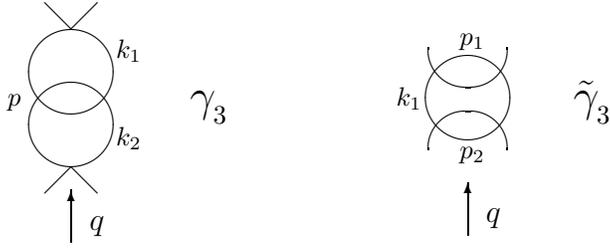}
\caption{\lb{jetp-3l}\bf Three-loop diagrams}
\end{center}
\end{figure}

The second example is the three-loop diagram $\g_3$ shown in figure 2.
Let $k_1$ be the loop momentum in the upper big loop,
$k_2$ be the loop momentum in the lower big loop,
and $p$ be the loop momentum in one-loop diagram in the center.
There are two regions that give contributions to the leading
logarithms:
$k_1<k_2<p$ and $k_2<k_1<p$.
The corresponding integral for the first region is
\be
I=\int_{q}^{\Ld}d\ln\frac{k_1^2}{\Ld^2}
\int_{k_1}^{\Ld}d\ln\frac{k_2^2}{\Ld^2}
\int_{k_2}^{\Ld}d\ln\frac{p^2}{\Ld^2}
=\int_0^\eta dy \frac{1}{2}y^2
=\frac{1}{3!}\eta^3,
\ee
as before $\eta=\ln\frac{\Ld^2}{q^2}$.
For the second region the answer is the same.
The final answer for the leading logarithm is the sum over these
regions
\be
F(\g_3)=\frac{1}{3}\eta^3.
\ee

The last example is the diagram $\td{\g}_3$ shown in figure 2.
Let $p_1,\;p_2$ be the loop momenta in one-loop diagrams,
$k_2$ be the loop momentum in the big loop  and
$q$ be the external momentum.
The leading logarithm comes from the region $p_1,\:p_2>k$,
consequently the integral has the form
\be
F(\td{\g}_3)=
\int_{q}^{\Ld}d\ln\frac{k^2}{\Ld^2}
\int_{k}^{\Ld}d\ln\frac{p_1^2}{\Ld^2}
\int_{k}^{\Ld}d\ln\frac{p_2^2}{\Ld^2}
=\int_0^\eta dy y^2
=\frac{1}{3}\eta^3.
\ee

In general case there is a correspondence between the maximal rooted trees
of divergent subgraphs and integrals that give a contribution to the
leading logarithms.
The vertices of the tree correspond to subgraphs of the graph.
The root corresponds to the graph itself.
If the root is on the top of the tree and all the edges
go downwards,
then an edge between two vertices of the tree shows
that the subgraph for the lower vertex belongs
to the subgraph for the upper vertex.
Since the tree is maximal, then each subgraph has only one loop more
than the maximal sub-subgraph of this subgraph.
Consequently each vertex has a corresponding loop momentum.
If $\g_1\subset\g_2$, then the corresponding loop momentum $p_1<p_2$.
The variable corresponding to the root, i.e. to the graph itself,
belongs to the interval
$(q,\Ld)$, where $q$ is a typical external momentum.
Each integration yields one power of the logarithm.
Thus the integral for a $k$-loop graph $\g_k$ has the form
\be
I^{(k)}=\int_0^{\td{y}} dy y^{k-1}=\frac{1}{k}\td{y}^k,
\ee
where $\td{y}=\ln\frac{\Ld^2}{\td{p}^2}$ is the external for
$\g_{k}$ momentum.
The integrand $y^{k-1}$ comes from the integration over subgraphs of
$\g_k$.
We see that each integration gives the contribution proportional to
$1/k$, where $k$ is the number of loops in the corresponding graph.
Thus the $n$-ple integral corresponding to the rooted tree $T_\g$
equals
\be
I(T_\g)=\frac{1}{T^!_\g}\eta^n,
\ee
where $T^!_\g$ is the product of numbers of loops in the subgraphs of
$\g$.
The final answer for the leading logarithm is
\be
F(\g_n)=\sum_{T^!_{\g_n}}\frac{1}{T^!_{\g_n}}
\left(\ln\frac{\Ld^2}{q^2}\right)^n.
\ee
We see that the coefficient in front of the leading logarithm
coincides with the coefficient in formula (\ref{coef1}).

So far we have considered only the one-loop beta function and the
leading logarithms.
In general case the Callan-Symanzik equation for the four-point
function has the form
\be\lb{dup}
\p_\tau F+\hat{\bt}\circ F-2\hat{\g}\circ F=0,
\ee
where the operator $\hat{\g}$ comes from the renormalizations of the
fields.
This operator acts by insertion of 1PI diagrams with two external
edges.
The solution of (\ref{dup}) may be written in the form
\be\lb{sp}
F=e^{-\tau\hat{\bt}+2\tau\hat{\g}}F_0.
\ee
The function $F_0$ is the initial condition for the
differential equation (\ref{dup}).
If $\bt$ and $\g$ functions are known,
then, using equation (\ref{sp}), one can find the answer for any Feynman
integral that contributes to $F$.

\bigskip
\bigskip
{\bf\large Conclusion and outlook}

In the paper we study a generalization of RG equation following from the Hopf
algebra of graphs.
This equation is used to find the coefficients in front of the leading
logarithms for individual Feynman integrals.
The derivation is based on the existence of the Hopf algebra and doesn't need
any particular properties of $\vp^4$ theory.
Consequently the method should apply in any theory
where the Hopf algebra exists.
The main problem is to find the corresponding Hopf algebra.
For example, in Yang-Mills theories this algebra should be consistent
with the gauge invariance and with the matrix structure.
The Hopf algebra consistent with the gauge invariance in abelian gauge
theories was found in the work
\cite{volovich}.
Whereas the problem of existence
of the Hopf algebra in matrix theories is solved in
\cite{mhopf}.
Thus it is plausible that the Hopf algebra can be defined in
Yang-Mills theories.
More bold hopes are connected with the possibility of application of
the Hopf algebra in non-euclidian mutli-loop calculations
\cite{smirnov}.

\bigskip
\bigskip
The author is indebt to A.Yu.Morozov for valuable discussions.
The work is supported by RFBR grant 03-02-17373.


\begin{thebibliography}{100}


 \bibitem{BSh}
N.N.Bogoliubov, D.V.Shirkov,
Introduction to the Theory of Quantized Fields,
3rd ed., Wiley-Interscience, 1980.


 \bibitem{Col}
J.Collins, Renormalization, Cambridge Univercity Press, Cambridge,
1984.


 \bibitem{chetyr:ll}
K.~G.~Chetyrkin,
Nuovo Cim., {\bf 103A}, 1653 (1990).



\bibitem{CK1}
 A.Connes, D.Kreimer;
 Commun.Math.Phys. {\bf 210}, 249-273 (2000);  hep-th/9912092.

 \bibitem{CK2}
 A.Connes, D.Kreimer;
 Commun.Math.Phys. {\bf 216}, 215-241 (2001); hep-th/0003188.


 \bibitem{CM}
A.Connes, H.Moscovici,
 Commun.Math.Phys. {\bf 198}, 199 (1998); math.dg/9806109.

 \bibitem{GMS}
 A.Gerasimov, A.Morozov, K.Selivanov,
 Int.J.Mod.Phys.A16:1531-1558,2001,
 hep-th/0005053

\bibitem{Kreimer:treef1}
D.~Kreimer,
Adv.\ Theor.\ Math.\ Phys.\  {\bf 3}, 3 (2000)
[Adv.\ Theor.\ Math.\ Phys.\  {\bf 3}, 627 (1999)]
hep-th/9901099.


\bibitem{mllrg}
D.~Malyshev,
Phys.\ Lett.\ B {\bf 578}, 231 (2004),
hep-th/0307301.


\bibitem{mllnrg}
D.~Malyshev,
hep-th/0402074.

\bibitem{polyakov}
A.M. Polyakov,
Sov. Phys. JETP {\bf 30}, 151 (1970).

\bibitem{volovich}
I.V.Volovich and D.V.Prohorenko,
Proceedings of the Steklov Mathematical Institute,
{\bf 147}, 166 (2004).
(In Russian).


 \bibitem{mhopf}
D.Malyshev;
JHEP {\bf 0205}, 013 (2002); hep-th/0112146.

\bibitem{smirnov}
V.~A.~Smirnov and E.~R.~Rakhmetov,
Theor.\ Math.\ Phys.\  {\bf 120}, 870 (1999)
[Teor.\ Mat.\ Fiz.\  {\bf 120}, 64 (1999)]
[arXiv:hep-ph/9812529].
%
V.~A.~Smirnov,
Phys.\ Lett.\ B {\bf 465}, 226 (1999)
[arXiv:hep-ph/9907471].



 \end{thebibliography}
\end{document}